\documentclass[aps,prl,reprint,superscriptaddress]{revtex4-1}

\usepackage{amsmath}
\usepackage{upgreek}
\usepackage{graphicx}

\begin{document}

% Paper title
\title{Pentamode metamaterials with independently tailored bulk modulus and mass density}

% Authors and affiliations
\author{Muamer~Kadic}\email{muamer.kadic@kit.edu}
\affiliation{Institute of Nanotechnology, Karlsruhe Institute of Technology (KIT), 76128 Karlsruhe, Germany}
\author{Tiemo~B\"{u}ckmann}
\author{Robert~Schittny}
\affiliation{Institute of Applied Physics, Karlsruhe Institute of Technology (KIT), 76128 Karlsruhe,Germany}
\author{Peter Gumbsch}
\affiliation{Institute for Applied Materials (IAM), Karlsruhe Institute of Technology (KIT), 76128 Karlsruhe, Germany}
\author{Martin~Wegener}
\affiliation{Institute of Applied Physics, Karlsruhe Institute of Technology (KIT), 76128 Karlsruhe,Germany}
\affiliation{Institute of Nanotechnology, Karlsruhe Institute of Technology (KIT), 76128 Karlsruhe, Germany}

\date{\today}

\begin{abstract}
We propose a class of linear elastic three-dimensional metamaterials for which the effective parameters bulk modulus and mass density can be adjusted independently over a large range---which is not possible for ordinary materials. First, we systematically evaluate the static mechanical properties and the phonon dispersion relations. We show that the two are quantitatively consistent in the long-wavelength limit. To demonstrate the feasibility, corresponding fabricated polymer microstructures are presented. Finally, we discuss calculations for laminates composed of alternating layers of two different metamaterials with equal bulk modulus yet different mass density. This leads to metamaterials with effectively anisotropic uniaxial dynamic mass density tensors. 
\end{abstract}

\maketitle

The mechanical properties mass density $\rho$ and bulk modulus $B$ (the inverse of the compressibility) are strongly correlated for ordinary substances. For example, an ideal gas held at temperature $T$ follows the universal relation $B/\rho=T \times \text{const.}$ \cite{Nag2005,Chocano2012}. There is no such strict relation for solids, but one gets a rough general correlation between the Young's €™modulus and $\rho^n$ ($1 \leq n \leq 3$) in the so-called Ashby plot \cite{Ashby2010,Zheng2014}. For constant Poisson's €ratio, this correlation translates to $B/\rho^n \approx \text{const.}$. Recently, ultralight-weight three-dimensional mechanical microlattices have been investigated by different groups \cite{Kistler1987,Tillotson1992,Schaedler2011,Mecklenburg2012,Yang2013,Bauer2014,Zheng2014}. Herein, ideally, one aims at reasonably small compressibility, {\it i.e.}, at reasonably large $B$ for small $\rho$. To another extreme, one may also want materials with large $\rho$ yet small $B$.

For example, in the context of coordinate-transformation mechanics \cite{Milton2006,Norris2008,Brun2009,Olsson2011,Norris2012a,Norris2013d,Craster2013,Diatta2014}, the ability to independently adjust $B$ and $\rho$ of a mechanical material at each point in space is crucial. For general cloaking structures, one would even like to tailor inhomogeneous anisotropies \cite{Scandrett2010,Cipolla2011}. This is in analogy to transformation optics \cite{Pendry2006,Leonhardt2006} based on the Maxwell equations, where the inverse electric permittivity $\epsilon^{-1}$ (tensor) is the counterpart of the bulk modulus $B$ (tensor) and the magnetic permeability $\mu$ (tensor) is the counterpart of the mass density $\rho$ (tensor) in mechanics \cite{Norris2008,Cummer2008}. Here, we have tacitly assumed that the shear modulus is zero or at least small. Otherwise, there is no such simple correspondence between mechanics and electromagnetism \cite{Milton2006}.

\begin{figure}
\includegraphics[scale=1]{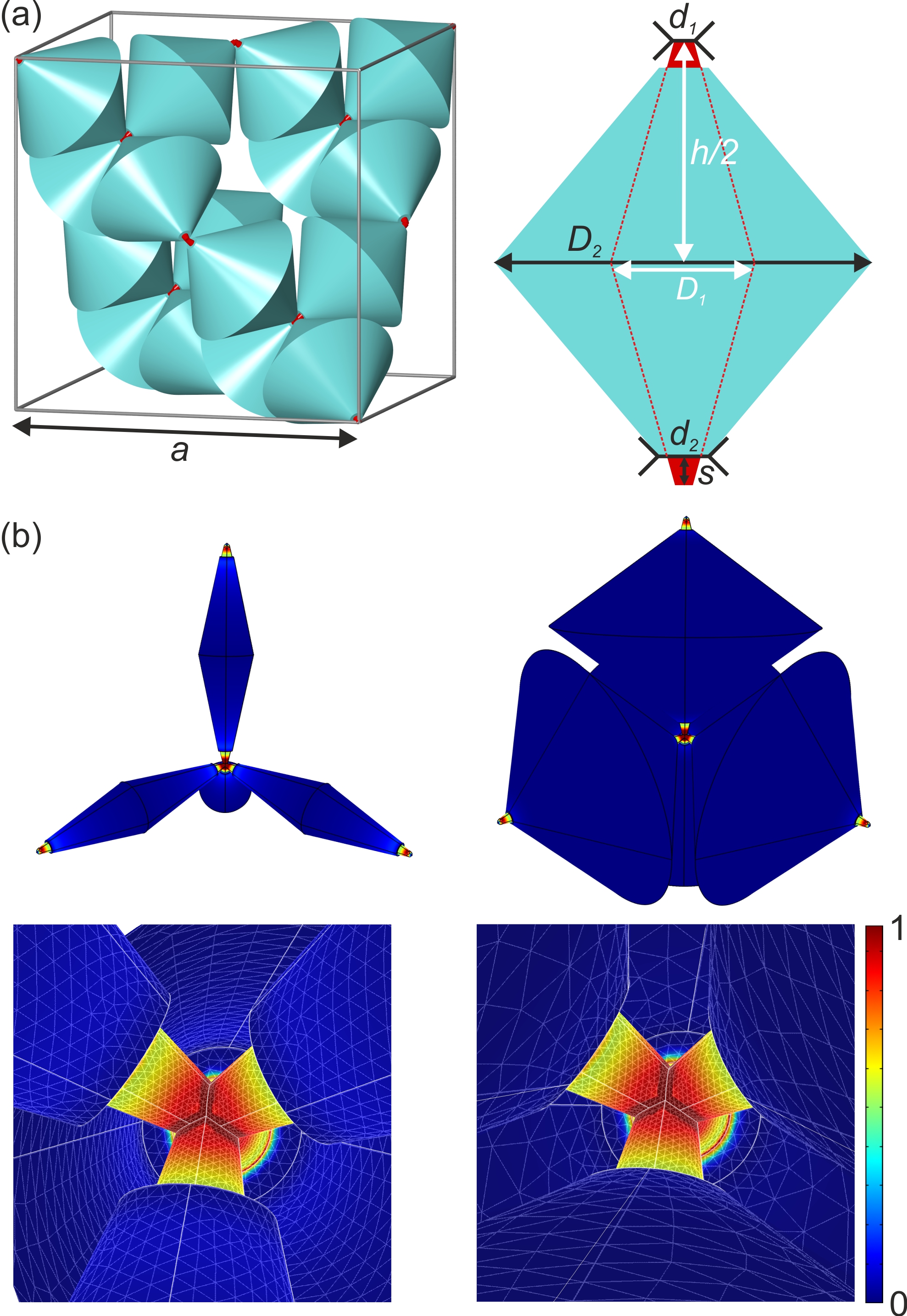}
\caption{(a) Illustration of the metamaterial structure. The red and blue parts refer to the identical constituent material, they are only colored for illustration. The parameters and especially the important diameters $d_1$ and $D_2$ as well as the lattice constant $a$ of the face-centered-cubic (fcc) lattce are defined. Due to the diamond lattice of connection points, the double-cone length $h$ is given by $h=a \sqrt{3}/4$. The ratio $d_1/a$ determines the effective bulk modulus $B/B_0$, the ratio $D_2/a$ determines the effective relative static mass density $\rho/\rho_0$, which is equal the volume filling fraction $f$. Throughout this paper, we fix the parameters $s/h=0.05$, $D_1/a=0.12$, and $d_2/a=0.04$. (b) This decoupling between $B$ and $\rho$ becomes possible due to the concentration of the stress to the immediate tip regions. Thus, the other stress-free regions are irrelevant for the bulk modulus. In turn, the tip regions are irrelevant for the mass density. Two examples of von Mises stress \cite{Mises1913} fields are depicted on a normalized false-color scale. The magnified views in the lowest row show the concentration of the normalised von Mises stess in the connection regions. The mesh used for the numerical calculations is overlaid.}
\label{fig1}
\end{figure}

How can we obtain the aimed-at independent control of $B$ and $\rho$ with an artificial crystal? We start from three-dimensional pentamode metamaterials \cite{Milton1995,Sigmund1995, Kadic2012,Kadic2013,Schittny2013,Mejica2013,Spadoni2014}, the counterpart of bimode metamaterials in two dimensions \cite{Layman2013}. These artificial materials have an effective shear modulus $G$ that is orders of magnitude smaller than their effective bulk modulus, {\it i.e.}, their effective Poisson's ratio approaches 0.5 from below. This can be achieved \cite{Kadic2012,Schittny2013} by a lattice of needle-like objects, the conical tips of which touch each other on a diamond lattice with corresponding face-centered cubic (fcc) lattice constant $a$ \cite{Kadic2012}. The key aspect for the present paper is the following: Upon exerting a hydrostatic pressure onto the pentamode structure, the stress field is essentially concentrated to the tip touching regions and their immediate surroundings, see Fig.~\ref{fig1} (the underlying numerical calculations will be specified below). The stress is negligible in the other parts. This means that the other parts do not significantly influence the elastic properties at all. We can thus add mass to these other regions to independently tailor the mass density $\rho$ without affecting $B$. The volume filling fraction $f$ of fabricated microscopic \cite{Kadic2012} and macroscopic \cite{Schittny2013} ordinary pentamode metamaterials has been as small as a few per cent and yet much smaller values of $f$ are conceptually possible. This leaves plenty of room towards the ultimate maximum of $f=100\%$. Thus, the mass density can be varied over an order of magnitude with present fabrication technology and conceptually even much more---while fixing the bulk modulus $B$. To be able to efficiently fill the volume and keep the overall structure simple at the same time, we have chosen the regions outside of the conical touching tips to be conical in shape as well. Their large diameter is $D_2$. Importantly, when changing $D_2$, the shape of the tips does not change at all, leading to a decoupling of the tips from the rest (see Fig.~1). The tips are illustrated in red, whereas the rest is highlighted in blue---although both are made from the identical constituent material.

We have shown previously \cite{Kadic2012} that the isotropic bulk modulus $B$ of ordinary pentamode metamaterials is simply proportional to the small diameter $d_1$ of the touching region (see Fig.~\ref{fig1}), provided that $d_1/a \ll 1$ and $D_2/d_1>1$. Varying the ratio $d_1/a$, and hence $B$, while maintaining the condition $d_1/a \ll 1$ for fixed $a$, seems possible over an order of magnitude with present technology. Again, conceptually, yet larger ranges are possible. In the same limit $d_1/a\ll 1,$ the shear modulus $G$ is negligible, {\it i.e.}, $G/B \ll 1 $ \cite{Kadic2012}. 

To summarize, for sufficiently small values of the small connection diameter, {\it i.e.}, for $d_1/a \ll 1 $, the $d_1/a$ ratio determines the ratio of the effective isotropic bulk modulus and the isotropic constituent material bulk modulus, $B/B_0$. The big diameter with respect to the lattice constant, $D_2/a$, mainly determines the volume filling fraction $f \in [0\%,100\%]$ and hence the effective static mass density $\rho$ via $f=\rho/\rho_0$, with the mass density of the constituent material $\rho_0$.
% % % % % % % % % % % % % % %
% % % % % % % % % % % % % % %
% % % % % % % % % % % % % % %
% % % % % % % % % % % % % % %
% % % % % % % % % % % % % % %
% % % % % % % % % % % % % % %
% % % % % % % % % % % % % % %
% % % % % % % % % % % % % % %

\begin{figure}
 \includegraphics[scale=0.8]{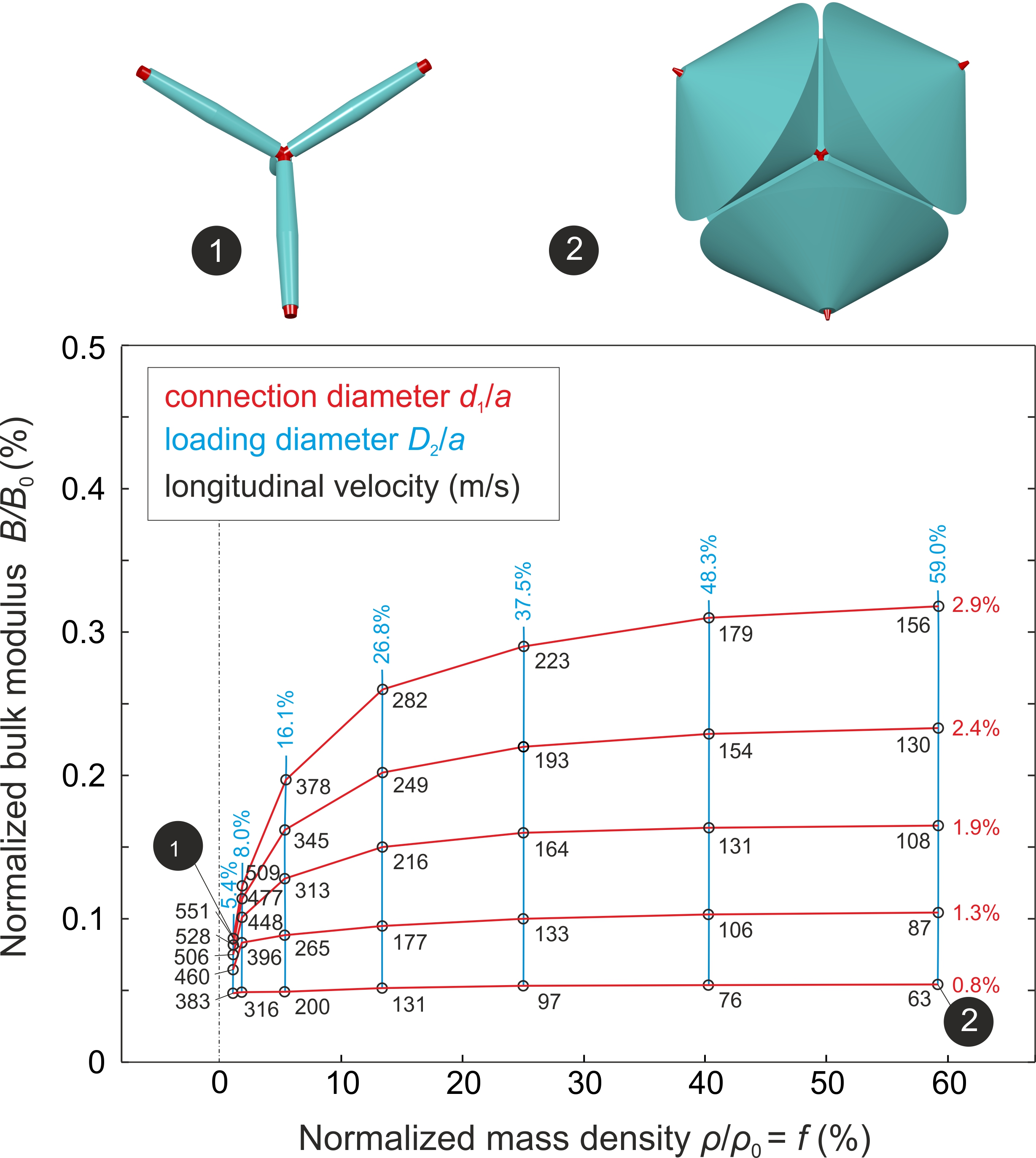}
 \caption{Calculated metamaterial properties in the $B \rho$ plane. The red lines connect points with constant ratio $d_1/a$, the blue lines points with constant $D_2/a$. Obviously, a large range of bulk moduli $B/B_0$ and mass densities $\rho/\rho_0=f$ can be accessed. Here the quantities with index ``0'' refer to the constituent bulk material. The resulting effective sound velocities $c=B/\rho$ are also given at the data points. They refer to a typical polymer as constituent material. All effective sound velocities $c=\sqrt{B/\rho}$ agree to within numerical accuracy ($<1\%$) with the respective velocities $c=\omega/|\vec{k}|$ derived from the phonon band structures (see Fig.~\ref{fig3}). }
 \label{fig2}
 \end{figure}

Figure~\ref{fig2}(a) illustrates the properties of the considered metamaterials in the $B \rho$ plane. Two selected extreme unit cells are shown for illustration. The red curves correspond to different values of constant ratio $d_1/a$, the blue curves to different fixed ratios $D_2/a$. We emphasize once again that the decoupling becomes strict in the limit $d_1/a \to 0$ (in which case the blue and red curves should all be horizontal and vertical straight lines, respectively), but one can see from Fig.~\ref{fig2} that the decoupling already works approximately for ratios $d_1/a$ in the range of just a few per cent. The calculations for the effective bulk modulus $B$ have been performed by numerically solving the continuum-mechanics equations using the commercial software package \textsc{COMSOL} Multiphysics (\textsc{MUMPS} solver, $3\times10^5$ degrees of freedom) and by applying a hydrostatic pressure from all sides of a structure composed of one extended fcc unit cell as in \cite{Bueckmann2014a}. For the constituent material, we take typical polymer parameters, {\it i.e.}, a bulk modulus of $B_0=5\times10^9\,\rm Pa$, a mass density of $\rho_0=1190\,\rm kg\,m^{-3}$, and a Poisson's ratio of $\nu_0=0.4$. As discussed previously \cite{Kadic2012}, the latter is not important at all and the others can easily be scaled (see below). The calculations for the effective mass density $\rho$ are simply based on the polymer volume filling fraction.
%, which could be obtained analytically but has actually been taken from the \textsc{COMSOL} Multiphysics calculations as well.
% % % % % % % % % % % % % % %
% % % % % % % % % % % % % % %
% % % % % % % % % % % % % % %
% % % % % % % % % % % % % % %
 % % % % % % % % % % % % % % %
 % % % % % % % % % % % % % % %
 % % % % % % % % % % % % % % %
 % % % % % % % % % % % % % % %
Within the framework of continuum mechanics considering pentamode metamaterials ({\it i.e.}, negligible shear resistance), the phase velocity of longitudinally polarized compression waves is given by $c=\sqrt{B/\rho}$ (= sound velocity). The acoustic wave impedance is $Z=\sqrt{B \rho}$. 

To connect static continuum mechanics to elastic wave propagation, we have also calculated the phonon band structures $\omega(\vec{k})$ of the modified pentamode metamaterials systematically as a function of  the two design parameters $d_1/a$ and $D_2/a$. We consider the polymer structure in vacuum and again use \textsc{COMSOL} Multiphysics (\textsc{MUMPS} solver, Floquet-Bloch periodic boundary conditions imposed onto the primitive unit cell, about $6\times10^5$ degrees of freedom, compare \cite{Martin2012}). Examples are exhibited in Fig~3. We only depict wave vectors $\vec{k}$ along the $\Gamma$K direction because we have previously shown that the pentamode dispersion relation is isotropic at small $d_1/a$ \cite{Martin2012} although the structure itself has only cubic symmetry. Obviously, static continuum mechanics (red straight line) and dynamic wave propagation (black points) are quantitatively consistent in regard to the effective longitudinal phase velocities $c$ within the long-wavelength or effective-medium limit. More derived phase velocities are presented as the numbers at the data points in Fig.~\ref{fig2}. For all combinations of $d_1/a$ and $D_2/a$, the longitudinal phase velocity derived from the band structure, $c=\omega /|\vec{k}|$,  and that derived from continuum mechanics, $c=\sqrt{B/\rho}$, agree to within less than $1\%$ relative difference. This agreement for all parameters strongly suggests that, for the  metamaterial, the effective dynamic mass density is closely similar to the static mass density and, likewise, the effective dynamic bulk modulus is closely similar to the static bulk modulus. In general, the dynamic mass density can be quite different from the static mass density \cite{Mei2007}.
 
The slower transversely polarized shear modes connected to soft shear springs as well as the flat ``deaf'' bands, which correspond to localized vibrations of the masses, are de-emphasized in light gray. As to be expected, these flat bands move downwards in frequency in the calculations with increasing mass ({\it i.e.}, larger $D_2/a$) and/or with decreasing spring constant ({\it i.e.}, smaller $d_1/a$). In an intentionally inhomogeneous structure, one might couple to these modes though. This could be avoided by choosing operation frequencies outside of these flat bands.

\begin{figure}
\includegraphics[scale=1.8]{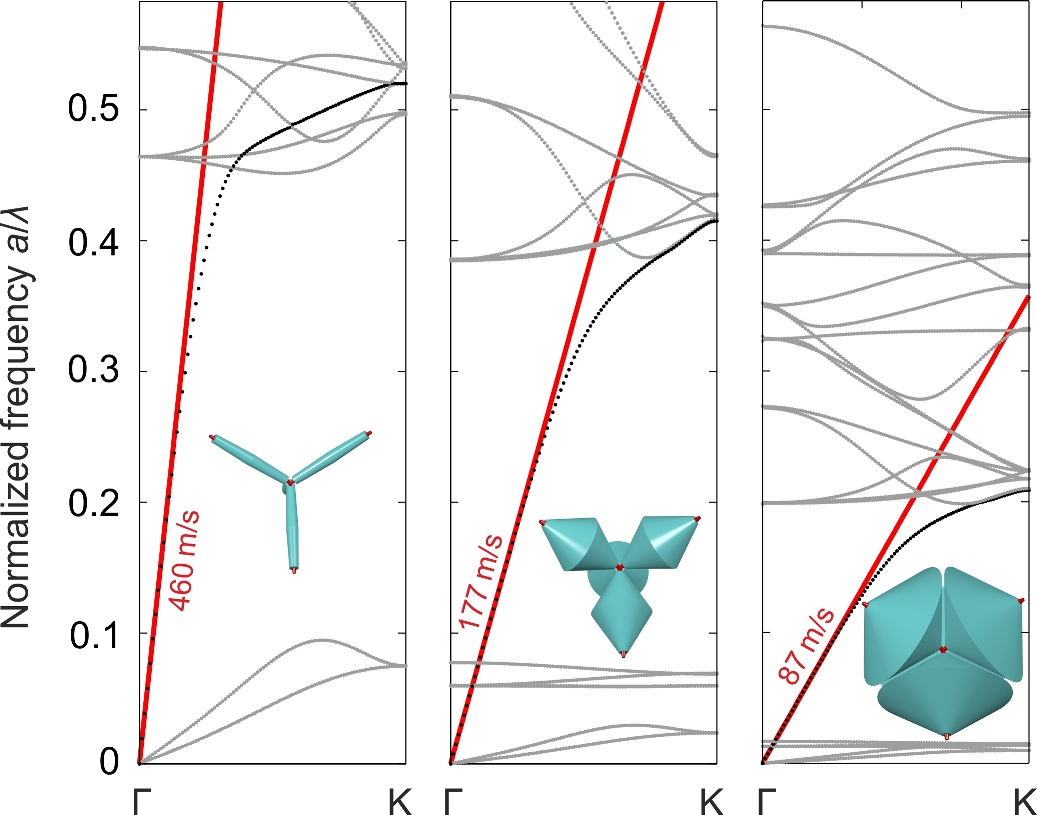}
\caption{Calculated phonon band structures $\omega(\vec{k})$ for three different parameter sets. $d_1/a=1.34\%$ (a) $D_2/a=5.4\%$, (b) $D_2/a=26.8\%$, and (c) $D_2/a=59.0\%$; all other parameters are given in the main text. The longitudinal branch is plotted in black, all other modes are de-emphasized in light gray. Only the $\rm \Gamma K$ (or (111) or along space diagonal) direction is depicted because the dispersion relation for the relevant longitudinal branch is very nearly isotropic (the transversely polarized shear modes are not isotropic though). The velocities $c=\omega/|\vec{k}|$ computed for many other parameters combination are given as numbers in Fig.~2. The red straight line is the dispersion relation corresponding to continuum mechanics with $\omega=|\vec{k}| \sqrt{B/\rho}$ with the values for $B$ and $\rho$ as shown in Fig.~2.}
\label{fig3}
\end{figure}

In passing, we emphasise again the scalability of our results: The dimensionless normalized frequencies $a/\lambda$ on the horizontal axes of Fig.~\ref{fig3} refer to a wavelength $\lambda$ in standard air (at 20 degrees Celsius) with an air velocity of sound $c_{\rm air}=343\,\rm m/s$. The absolute frequency $\omega/(2 \pi)$ in units of Hz results from the dispersion relation $\omega /(2\pi)\lambda=c_{\rm air}$. Comparing to air is meaningful because one may eventually couple such structures to airborne sound. In any case, it is simple to scale the wavelength to other media. For example, for a lattice constant of $a=40\,\rm \mu m$ and for the above polymer constituent material parameters, a normalized frequency of $a/\lambda=0.5$ corresponds to an absolute frequency of $4.3\,\rm MHz$. On the basis of the continuum-mechanics equations, it is also straightforward to scale our results to metamaterials made from other constituent materials. For a constituent material bulk modulus $\tilde{B}_0$ and mass density $\tilde{\rho}_0$, instead of our above choices $B_0$ and $\rho_0$, one gets
$\omega \to \omega \times \sqrt{\frac{\tilde{B}_0 \rho_0}{B_0 \tilde{\rho}_0}}$.
 % % % % % % % % % % % % % % %
 % % % % % % % % % % % % % % %
 % % % % % % % % % % % % % % %
 % % % % % % % % % % % % % % %
 % % % % % % % % % % % % % % %
 % % % % % % % % % % % % % % %
 % % % % % % % % % % % % % % %
 % % % % % % % % % % % % % % %
 
\begin{figure}
\includegraphics[scale=1.5]{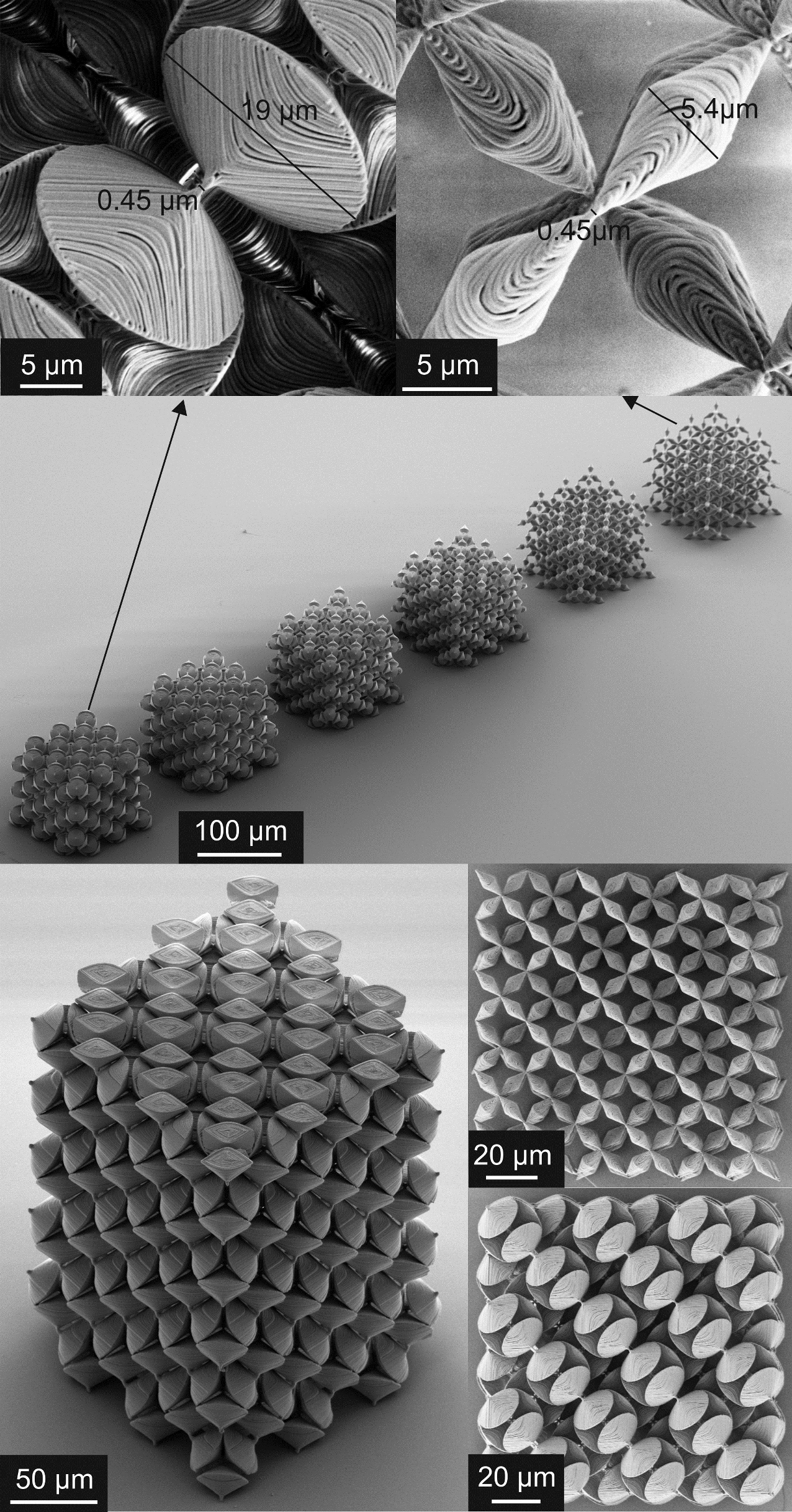}
\caption{Electron micrographs of selected fabricated metamaterial samples with fcc lattice constant $a=40\,\rm \mu m$. These polymer structures have been made by three-dimensional dip-in galvo-scanner-based laser lithography and demonstrate the feasibility of the concept of modified pentamode metamaterials.}
\label{fig4}
\end{figure}

Are such complex modified pentamode metamaterials with fine features at the connections and large masses in between experimentally feasible with current technology? Would the connections collapse under the large weight? To address these questions, we have fabricated polymer-based test samples with a lattice constant of $a=40\, \rm \mu m$ using state-of-the-art three-dimensional dip-in galvo-scanner-based optical laser lithography (Nanoscribe GmbH, Photonic Professional GT). Details of this technology can be found in \cite{Bueckmann2014,Bueckmann2014a}. Electron micrographs of fabricated metamaterial samples are depicted in Fig.~\ref{fig4}. Obviously, the structures are of high quality and the necessary aspect ratios and extreme three-dimensional motifs are possible. However, within current state-of-the-art, we can only fabricate up to around $10 \times 10 \times 10=10^3$ extended fcc unit cells \cite{Bueckmann2014a} on a timescale of ten hours. For meaningful wave measurements that are not dominated by edge effects, one rather needs around $100 \times 100 \times 100=10^6$ extended fcc unit cells or more. These would require excessive writing times on the scale of half a year. Such structures may come into reach though in some years as three-dimensional micro-printing gets yet faster.
 % % % % % % % % % % % % % % %
 % % % % % % % % % % % % % % %
 % % % % % % % % % % % % % % %
 % % % % % % % % % % % % % % %
 % % % % % % % % % % % % % % %
 % % % % % % % % % % % % % % %
 % % % % % % % % % % % % % % %
 % % % % % % % % % % % % % % %

Next, as a conceptual application example, we consider laminates made of two different modified pentamode metamaterials to obtain effectively anisotropic dynamic mass density tensors. It is well known that laminates composed of alternating layers of locally isotropic materials can lead to effectively anisotropic metamaterial behavior \cite{Kadic2013b}. A simple case is a stack of alternating good and bad electrically conducting layers that leads to large conductivity for currents flowing in the plane and small conductivity normal to the plane \cite{Milton2002,Kadic2013b}. We aim at a similar anisotropic behavior for the phase velocity of longitudinal waves propagating in the plane of the layers and normal to them, respectively. Using the metamaterials introduced above, we can now consider a special situation in which all the layers have the same bulk modulus, but where the mass density is alternating between high and low with a contrast of, e.g., about a factor of ten. Correspondingly, we expect the wave propagation anisotropy to be due to the mass density tensor anisotropy only.

\begin{figure*}[ht!]
\includegraphics[scale=1]{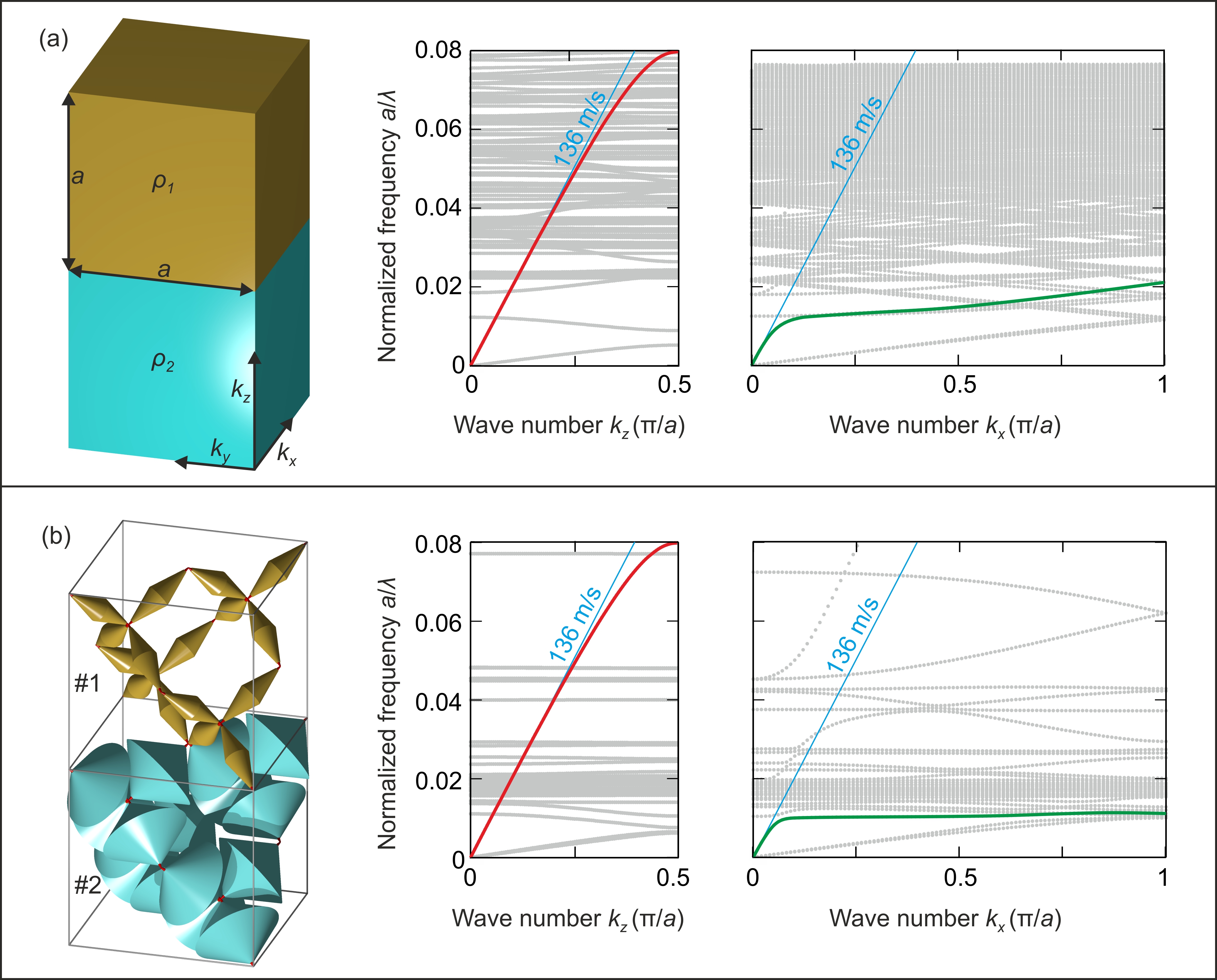}
\caption{(Two-column) (a) Left: Illustration of a laminate metamaterial composed of alternating bulk layers with identical bulk modulus $B_1=B_2=5.9\times10^6\,\rm Pa$, but different mass densities $\rho_1=47.4\,\rm kg\,m^{-3}$ and $\rho_2=588\,\rm kg\,m^{-3}$. Middle and right: corresponding calculated band structures for wave propagation normal to the laminate layers (red, $z$-direction) and in the layer plane (green, in $xy$-plane), respectively. The relevant longitudinal (``acoustic'') modes are highlighted in color. (b) Same, but for a metamaterial microstructure as shown on the left-hand side. To obtain the same effective bulk modulus yet different mass densities for the two individual laminate layers, we have not only varied $D_2$ (which mainly determines $\rho$) but have also slightly adjusted $d_1$ (which mainly determines $B$), {\it i.e.}, we have chosen $d_1/a=1.88\%$ ($d_1/a=1.34\%$) and $D_2/a=13.4\%$ ($D_2/a=53.6\%$) for layer $\#$1($\#$2). All other parameters are fixed and as quoted in Fig.~\ref{fig1}.}
\label{fig5}
\end{figure*}

Figure~\ref{fig5} summarizes band structure calculations $\omega (\vec{k})$ for laminates composed of bulk layers for reference (panel (a)) and for the actual complex laminate metamaterial structure (panel (b)), respectively. Here, each laminate layer has a thickness of one extended fcc pentamode lattice constant $a$. This leads to the complex heterostructure illustrated in Fig.~\ref{fig4}(b). Slow acoustic bands emerging from the zone center corresponding to transverse waves connected to the small shear modulus as well as bands corresponding to modes with flat dispersion (roughly related to optical phonons in ordinary crystals with a two-atom basis) are again plotted in light gray to emphasize in color the more important longitudinal acoustic compression waves connected to the bulk modulus. In the true long-wavelength limit, we find an isotropic behavior for the phase velocity $c=\omega/|\vec{k}|$. This is expected for any type of structure because all masses oscillate in phase in the quasi-static limit \cite{Milton2002,Kadic2013b}. For higher frequencies, yet for wave vectors $\vec{k}$ still well separated from the edge of the first Brillouin zone (this means the effective-medium description is appropriate), the bands for wave propagation in the plane of the laminate layers and normal to them, respectively, separate (compare red and green curves in the middle). The normal ($z$-direction) mode crosses the various rather flat optical-phonon-like bands as a straight line, whereas the mode propagating in the plane of the layers ($xy$-plane) bends over. 

 % % % % % % % % % % % % % % %
 % % % % % % % % % % % % % % %
 % % % % % % % % % % % % % % %
 % % % % % % % % % % % % % % %
 % % % % % % % % % % % % % % %
 % % % % % % % % % % % % % % %
 % % % % % % % % % % % % % % %
 % % % % % % % % % % % % % % %
  
Intuitively, if the shear was actually zero, one would get a fast compression mode in the low-mass-density layer and an additional independent slower one in the high-mass-density layer. However, as the shear is small but not zero, the high-mass-density layer is pulled back by the shear with respect to the low-mass-density layer (and vice versa). This small shear force together with the large mass of the high-mass-density layer leads to a mass-and-spring system with low eigenfrequency. For excitation above this resonance frequency, this mass reacts with 180 degrees phase shift, which can be described as an effectively negative mass density \cite{Kadic2013b}. Altogether, one gets a different effective dynamic mass density for propagation in the laminate layer plane and perpendicular to it, respectively, {\it i.e.}, the scalar isotropic mass density of the metamaterial turns into an anisotropic mass density tensor for the laminate. One should be aware though that this anisotropy can be exploited only over a fairly small frequency region. This limitation holds true for any structure exhibiting a resonant dynamic mass density (tensor), because the mass density of an elastic solid always approaches the isotropic static case, $\rho_{\rm stat}=f \rho_0$ , in the true long-wavelength limit (see discussion above)---just like the magnetic permeability of electromagnetic metamaterials approaches $\mu=1$ in the static limit \cite{Pendry1999}. 

In conclusion, we have introduced modified pentamode metamaterials and laminate heterostructures made thereof. These structures provide enhanced flexibility for molding the flow of longitudinal acoustic phonons in mechanics similar to that for photons in magneto-dielectric metamaterials in optics. Corresponding polymeric three-dimensional unit cells can be fabricated with current state-of-the-art optical laser lithography.

\begin{acknowledgments}
We thank the Hector Fellow Academy for support. We also acknowledge support by the Deutsche Forschungsgemeinschaft (DFG), the State of Baden-W\"urttemberg, and the KIT via the DFG-Center for Functional Nanostructures (CFN) through projects A\,1.4 and A\,1.5 and via the Karlsruhe School of Optics $\&$ Photonics (KSOP).
\end{acknowledgments}
M.K. and T.B. contributed equally to this work.

%\bibliography{references}

%\begin{thebibliography}{99}

%\end{thebibliography}
\end{document}